\documentstyle{elsart}

\input{epsf}
\input amssym.def \input amssym

\begin{document}

\begin{frontmatter}

\title{Effect of Noise on Excursions To and Back From Infinity}
\author{Jeff Moehlis}\footnote{E-mail: jmoehlis@math.princeton.edu}
\address{Program in Applied and Computational Mathematics, Princeton
University, Princeton, NJ, 08544-1000}

\begin{abstract}
The effect of additive white noise on a model for bursting behavior in large 
aspect-ratio binary fluid convection is considered.  Such bursts are 
present in systems with nearly square symmetry and are the result of 
heteroclinic cycles involving infinite amplitude states created when the 
square symmetry is broken.  A combination of numerical results and
analytical arguments show how even a very small amount of
noise can have a very large effect on the amplitudes of successive bursts.  
Large enough noise can also affect the physical manifestations of the bursts.  
Finally, it is shown that related bursts may occur when white noise is added 
to the normal form equations for the Hopf bifurcation with exact square 
symmetry.

\end{abstract}



\end{frontmatter}

\section{Introduction}
In physical systems, noise is unavoidable.  Thus, it is important to study 
the effect of noise on models of physical systems.  It has been shown that 
some models are very sensitive to the presence of noise: for example, the 
model for the resonant interaction of three wave modes
considered in \cite{lyth93} shows that a tiny amount of noise can 
replace a bifurcation structure involving a period doubling cascade to 
chaos with a noisy periodic orbit which is attracting over a wide range 
of parameter values.  Noise can have an especially profound effect on 
models with structurally stable heteroclinic cycles.  These cycles involve
connections between unstable states, and the connections typically 
occur in invariant subspaces.  Even small amounts of noise can cause
the system to jump across these subspaces, and this can lead to very 
different dynamical behavior~\cite{ston99}.  If such a heteroclinic cycle 
also is {\it attracting}, the time 
spent in the neighborhood of the unstable states of the cycle increases 
without bound with increasing time.  This is not observed in physical systems 
and noise provides an explanation: the heteroclinic cycle becomes a 
``statistical limit cycle'' in which switching between the unstable states 
of the cycle occurs randomly in time but with a well-defined mean 
period~\cite{buss80,ston89,ston90}.
 
In~\cite{land96,moeh98}, a model of large aspect-ratio binary fluid 
convection was studied.  The model considers the competition between two 
nearly degenerate modes of opposite parity. The evolution equations 
correspond to a Hopf bifurcation with broken D$_4$ symmetry, where the 
``interchange'' symmetry between the two modes is weakly broken because 
of the large but finite aspect-ratio.  For open parameter regimes, 
periodic or irregular bursting with very large dynamic range can occur 
close to threshold.  The bursts are associated with nonattracting 
heteroclinic cycles involving solutions ``at infinity''~\cite{moeh98,moeh00}.
These cycles may be structurally stable or unstable.  Bursts occur when 
the system evolves toward an infinite amplitude state along its stable 
manifold, then gets kicked toward another infinite amplitude state with 
an unstable manifold which returns the system to finite amplitude.  This 
mechanism provides an explanation for bursts found experimentally for 
large aspect-ratio binary fluid convection~\cite{sull88}, and may also 
provide a deterministic explanation for the variability of the solar 
magnetic cycle~\cite{knob96b}.  A comparison of this to other mechanisms
for bursting is given in~\cite{knob99b,knob00}.  Related bursts can
occur due to the resonant temporal forcing of a system undergoing a
Hopf bifurcation with D$_4$ symmetry~\cite{moeh00b}.

When noise is added to the model considered in~\cite{land96,moeh98},
there will be a burst with larger amplitude if 
the noise kicks the system closer to the stable manifold of an infinite 
amplitude state, and a burst with smaller amplitude if the noise kicks it 
away.  Thus a state which bursts in the absence of noise is still expected 
to burst in the presence of noise, but the successive burst amplitudes and 
the time between successive bursts may change substantially.  The details
of the effect of noise on the bursting behavior are described in this paper.  
This includes the 
possibility that noise can affect the physical manifestations of the bursts.
Specifically, purely blinking and winking states in which successive bursts 
occur at opposite sides or the same side of the container, respectively, 
can be destroyed by sufficiently large noise.  Finally, it is shown that 
related bursts can occur even if the D$_4$ symmetry of the model is 
{\it not} broken, i.e., when noise is added to the normal form equations 
for the Hopf bifurcation with {\it exact} D$_4$ symmetry.  


\section{Evolution equations for the model}

For binary fluid convection, if the separation ratio
of the mixture is sufficiently negative then convection arises via a
Hopf bifurcation.  This is the case for the $^3$He/$^4$He mixture used
by Sullivan and Ahlers~\cite{sull88} in their experiment carried out in
the large aspect-ratio rectangular container
\[
D \equiv \left\{ X,Y,Z|-\frac{1}{2}\Gamma \le X \le \frac{1}{2} \Gamma,-\frac{1}{2}\Gamma_Y \le Y \le \frac{1}{2} \Gamma_Y,-\frac{1}{2} \le Z \le \frac{1}{2} \right\},
\]
with aspect-ratio $\Gamma = 34$, and $\Gamma_Y = 6.9$.  In this experiment,
it was observed that immediately above threshold convective heat
transport may take place in a sequence of irregular bursts of large dynamic
range despite constant heat input. 
A model for this experiment is considered in~\cite{land96,moeh98}.  The 
perturbation to the conduction state temperature profile is assumed to 
take the form
\[
\Theta(X,Y,Z,t) = \epsilon^{1/2} {\rm Re} \{z_+(t) f_+(X,Y,Z) + z_-(t) f_-(X,Y,Z)\} + {\cal O}(\epsilon),
\]
where $\epsilon \ll 1, f_\pm(-X,Y,Z) = \pm f_\pm(X,Y,Z)$.  Here $z_+$ are $z_-$
are the (complex) amplitudes of the first modes to lose stability which 
are respectively even and odd under the reflection $X \rightarrow -X$.  
In~\cite{land96}, evolution equations for $z_+$ and $z_-$ were derived using 
symmetry arguments, with the resulting equations describing a Hopf bifurcation 
with broken D$_4$ symmetry.  The applicability of this model to the
experiment is discussed further in~\cite{bati99}.  As shown 
in~\cite{land96,moeh98}, solutions of the evolution equations include bursts 
of very large dynamic range similar to those found in the experiments. 
The mechanism by which these bursts arise is discussed in detail 
in~\cite{moeh00}, and is summarized below.

In this paper we generalize the model considered in~\cite{land96,moeh98}
to include additive random forcing terms.  Specifically, we 
consider the formal equations
\begin{eqnarray}
\frac{d z_+}{dt} &=& (\lambda + \Delta \lambda + i (\omega + \Delta \omega)) z_+ + A (|z_+|^2 + |z_-|^2) z_+ \nonumber \\
&& + B |z_+|^2 z_+ + C \bar{z}_+ z_-^2 + \eta_1(t)+ i \eta_2(t) \label{z+} \\
\frac{d z_-}{dt} &=& (\lambda - \Delta \lambda + i (\omega - \Delta \omega)) z_- + A (|z_+|^2 + |z_-|^2) z_- \nonumber \\
&& + B |z_-|^2 z_- + C \bar{z}_- z_+^2 + \eta_3(t) + i \eta_4(t) \label{z-}
\end{eqnarray}
where the $\eta_i$'s represent real, independent Gaussian white noise random 
processes with the properties
\begin{equation}
\langle \eta_i (t) \rangle = 0, \qquad \langle \eta_i(t) \eta_j(t') \rangle = 2 {\cal D} \; \delta(t-t') \; \delta_{ij}, \qquad i,j=1,2,3,4.
\end{equation}
Here $z_+,z_-, A, B, C$ are complex, and 
$\lambda, \Delta \lambda, \omega, \Delta \omega, {\cal D}$ are real.
The quantity $\Delta \omega$ measures the difference in frequency between 
the two modes at onset, and $\Delta \lambda$ measures the difference in 
their linear growth rates.  The terms involving $\Delta \lambda$ and 
$\Delta \omega$ are called forced symmetry-breaking terms because they break 
the D$_4$ symmetry of the governing equations~\cite{moeh98}.  The new 
$\eta_i$ terms represent unavoidable random effects in the experiment.

We first summarize results for the case that there is no noise, i.e., when 
${\cal D} = 0$.  New coordinates $(\rho,\theta,\phi,\psi)$ are defined
according to
\[
z_+ = \rho^{-1/2} \cos(\theta/2) e^{i (\phi + \psi)/2}, \qquad z_- = \rho^{-1/2} \sin(\theta/2) e^{i (-\phi + \psi)/2},
\]
where without loss of generality $\theta \in [0,\pi]$, 
$\phi \in [-2 \pi,2 \pi)$, and $\psi \in [0,4 \pi)$.  Also introducing a 
new time $\tau$ defined by $d\tau/d t = 1/\rho$, equations~(\ref{z+},\ref{z-})
with ${\cal D}=0$ take the form
\begin{eqnarray}
\frac{d \rho}{d \tau} &=& -\rho [2 A_R + B_R (1 + \cos^2 \theta) 
+ C_R \sin^2 \theta \cos 2 \phi] 
- 2 (\lambda + \Delta \lambda \cos \theta) \rho^2, \label{rho} \\
\frac{d \theta}{d \tau} &=& \sin \theta [\cos \theta (-B_R + C_R \cos 2 \phi) 
- C_I \sin 2 \phi] - 2 \Delta \lambda \sin \theta \rho, \label{theta_rho} \\
\frac{d \phi}{d \tau} &=& \cos \theta (B_I - C_I \cos 2 \phi) 
- C_R \sin 2 \phi + 2 \Delta \omega \rho, \label{phi_rho} \\
\frac{d \psi}{d \tau} &=&  2 A_I + B_I + C_I \cos 2 \phi 
+ C_R \sin 2 \phi \cos \theta + 2 \omega \rho. \label{psi_rho}
\end{eqnarray}
Here $A = A_R + i A_I$, etc.  Let
\[
r = \frac{1}{\rho} = |z_+|^2 + |z_-|^2
\]
denote the amplitude of a solution.  As discussed in \cite{moeh98,moeh00}, 
equations (\ref{rho}-\ref{psi_rho}) have an invariant subspace $\Sigma$ 
at $\rho=0$ (corresponding
to infinite amplitude states) on which the equations are equivalent to those 
with $\Delta \lambda=\Delta \omega = 0$.  Since the variable $\psi$ 
decouples from the other variables, the 
dynamics on $\Sigma$ are two-dimensional and hence simple to 
analyze~\cite{swif88}.  Such an analysis (see ~\cite{moeh00}) allows us to 
conclude that, for this truncation, infinite amplitude periodic and 
quasiperiodic solutions exist.  Bursts are associated with heteroclinic 
cycles involving such infinite amplitude states.  
Reference~\cite{moeh00} shows that for such cycles to form, it is necessary
to have at least one subcritical and one supercritical solution branch
bifurcating from the trivial state with $\Delta \lambda=\Delta \omega = 0$;
also, the ``angular stability'' properties of the solutions must be
correct.  Reference~\cite{moeh00} also shows that despite their heteroclinic 
nature, the duration of the resulting bursts in the original time $t$ is in 
fact {\it finite}, i.e., the excursion to and return from infinity occur in 
{\it finite} time.
Of course, the infinite amplitude solutions and connections to 
them are of physical interest only insofar as they are responsible for the 
presence of nearby solutions which make visits to large but finite amplitude.
This type of bursting behavior persists for ${\cal D}=0$ even when higher 
order terms in equations (\ref{z+},\ref{z-}) are retained~\cite{moeh00}.
It should be emphasized that the noise is added on the original timescale 
$t$, not $\tau$ (see equations~(\ref{z+},\ref{z-})).

\section{Effect of noise on bursting behavior}
We now describe the effect of noise on the bursting behavior 
reported in~\cite{land96,moeh98}.  Coefficient values which give the 
appropriate sub- and supercriticality of traveling and standing wave 
branches in large aspect-ratio binary fluid convection systems
are considered~\cite{moeh98}:
\[
A = 1 - 1.5 i, \qquad B = -2.8 + 5 i, \qquad C = 1+i, 
\]
\[
\omega = 1, \qquad \Delta \lambda = 0.03, \qquad \Delta \omega = 0.02.
\]
For these coefficient values in the absence of noise, nonattracting, 
structurally stable heteroclinic cycles connecting infinite amplitude 
periodic orbits (hereafter, $u_\infty$ solutions) and infinite amplitude 
quasiperiodic orbits (hereafter, $qp_\infty$ solutions) exist over 
a range of $\lambda$ values.  These heteroclinic cycles are of 
Shil'nikov-Hopf type, and there are associated stable solutions which 
display bursting behavior~\cite{moeh98,moeh00}.

\begin{figure}[t]
\begin{center}
\leavevmode
\epsfbox{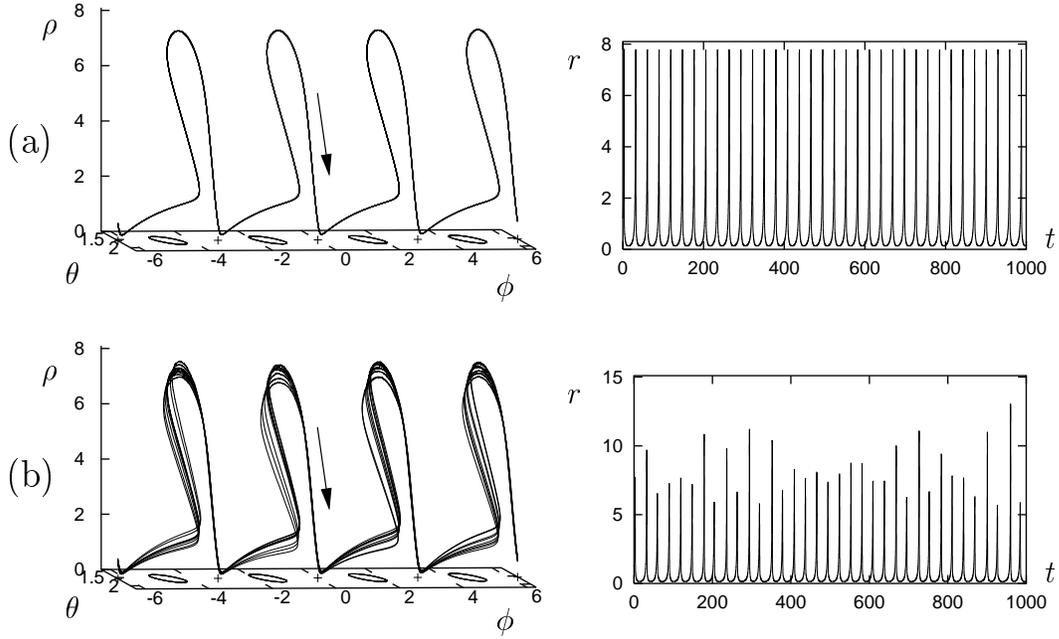}
\end{center}
\caption{
Bursts for $\lambda = 0.1$ and (a) ${\cal D} = 0$, (b) 
${\cal D} = 1 \times 10^{-7}$.  Clearly, even a very small amount of noise 
can have a very large effect on the burst amplitudes.  Fixed points (periodic
orbits) in these projections correspond to periodic orbits (quasiperiodic
orbits) for the full system.  In this and later figures, the $u_\infty$ 
solutions are denoted by $+$'s, and the $qp_\infty$ solutions appear as 
periodic orbits at $\rho=0$.
\label{blinking}}
\end{figure}

For definiteness, consider the stable quasiperiodic solution which is 
present in the absence of noise for $\lambda = 0.1$ (see 
Figure~\ref{blinking}(a); in the notation of~\cite{moeh00}, this is
a $w_e^1$ solution).  
The trajectory makes successive visits near {\it different} but 
symmetry-related $u_\infty$ solutions given by
$(\rho,\theta,\phi,\psi) = (0,\pi/2,m \pi,(2 A_I + B_I + C_I) \tau \; {\rm mod} \; 4 \pi)$,
$m = 0,\pm 1, \pm 2$.
The $u_\infty$ solutions are unstable within $\Sigma$, and
the trajectory gets kicked toward a $qp_\infty$ solution.  
The $qp_\infty$ solutions are unstable in the $\rho$ direction, so the 
trajectory returns to smaller amplitude (larger~$\rho$).  
Physically, this solution corresponds to a {\it blinking state} for the 
convection system because successive bursts occur at opposite sides of 
the container~\cite{moeh98}.  Figure~\ref{blinking}(b) shows the 
corresponding results for ${\cal D} = 1 \times 10^{-7}$.   Clearly, even 
this very small amount of noise can have a very large effect on the burst 
amplitudes.  These and other numerical results were obtained using a 
stochastic second-order Runge-Kutta algorithm~\cite{hone92}.  Unless 
otherwise indicated, the time step of integration is 
$\delta t = 1 \times 10^{-4}$.

\begin{figure}[t]
\begin{center}
\leavevmode
\epsfbox{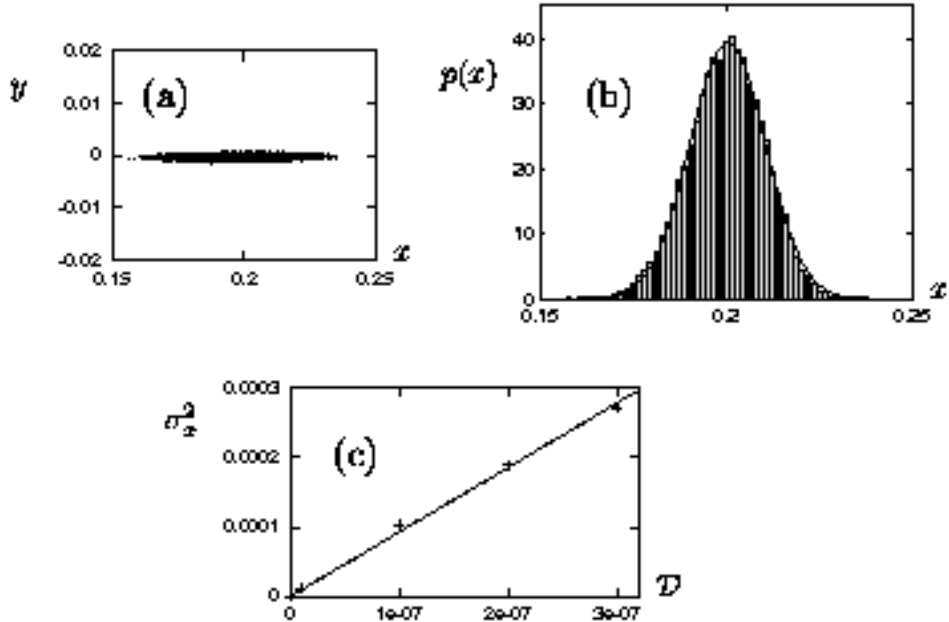}
\end{center}
\caption{
(a) Values of $(x,y)$ at which the trajectory intersects the surface $\rho=1$ 
with decreasing $\rho$ for ${\cal D} = 1 \times 10^{-7}$.  The stable 
manifold of the $u_\infty$ solutions intersects the surface $\rho=1$ 
approximately along the line $x=0$.  (b) Probability distribution function 
$p(x)$ at the surface $\rho=1$.  The histogram is from a long-time integration
of equations~(\ref{z+},\ref{z-}), and the solid line shows a fit to
equation~(\ref{pdf_x}).  (c) Variance $\sigma_x^2$ from fitting 
$p(x)$ to equation~(\ref{pdf_x}) for different noise strengths ${\cal D}$.
\label{incoming}}
\end{figure}

The results shown in Figure~\ref{blinking}(b) may be qualitatively understood 
by recognizing that there will be a burst with larger amplitude if the noise 
kicks the trajectory closer to the stable manifold of a $u_\infty$ solution, 
and a burst with smaller amplitude if the noise kicks it away.  This is 
elucidated by ignoring the uncoupled variable $\psi$, linearizing equations 
(\ref{rho}-\ref{phi_rho}) about the $u_\infty$ solutions, and defining 
coordinates $(\rho,x,y)$ to give the diagonal form
\begin{equation}
\frac{d x}{d \tau} = \lambda_u x, \qquad \frac{d \rho}{\d \tau} = -\lambda_s \rho, \qquad \frac{d y}{d \tau} = -\lambda_{ss} y,
\label{linearization}
\end{equation}
where $\lambda_u,\lambda_s,\lambda_{ss}>0$.  The stable manifold of the 
$u_\infty$ solutions then approximately intersects a plane of sufficiently 
small, constant $\rho$ along the line $x = 0$.  For the parameters 
under consideration, $\lambda_u = 0.0682$, $\lambda_s = 0.2$, and 
$\lambda_{ss} = 5.868$, and equation~(\ref{linearization}) is obtained
by using the coordinates approximately given by
\[
\left( \begin{array}{c} \rho \\ x \\ y \end{array} \right) = \left( \begin{array}{ccc} 1 & 0 & 0 \\ 0.269 & -0.730 & 0.706 \\ 0.00626 & 0.902 & 0.467 \end{array} \right) \left( \begin{array}{c} \rho \\ \theta-\pi/2 \\ \phi-m \pi \end{array} \right).
\]
Figure~\ref{incoming}(a) shows values at which the trajectory intersects 
the surface $\rho=1$ with decreasing $\rho$ for ${\cal D} = 1 \times 10^{-7}$.
This is interpreted as showing how the noise affects the proximity of the 
trajectory to the stable manifold of the $u_\infty$ solutions.  
The probability distribution function (pdf) of $x$ values at the surface 
$\rho=1$ is shown in Figure~\ref{incoming}(b).  By definition, $p(x) d x$ 
is the probability that the $x$ value at the surface $\rho=1$ is between 
$x$ and $x + d x$.  This is fit reasonably well by the Gaussian distribution
\begin{equation}
p(x) = \frac{1}{\sqrt{2 \pi \sigma_x^2}} \exp \left[ -(x - \bar{x})^2 / (2 \sigma_x^2) \right].
\label{pdf_x}
\end{equation}
Figure~\ref{incoming}(c) shows that the variances $\sigma_x^2$ of fitted
Gaussians vary linearly with the noise strength~${\cal D}$.  Over this
range $\bar{x} \approx 0.201 $, independent of ${\cal D}$.  The noise 
strengths have been restricted to the range shown in the figure because for 
larger ${\cal D}$ the noise can kick the trajectory close enough to the 
stable manifold of a $u_\infty$ solution that the long-time integration of 
equations~(\ref{z+},\ref{z-}) becomes numerically difficult.

\begin{figure}[t]
\begin{center}
\leavevmode
\epsfbox{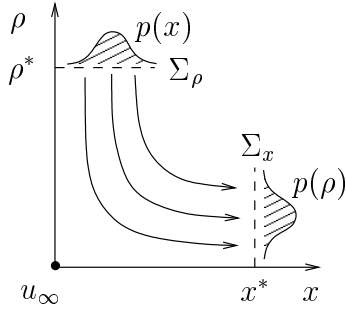}
\end{center}
\caption{
Setup for determining how the pdf of $x$ values at constant $\rho=\rho^*$
evolves into a pdf of $\rho$ values at constant $x=x^*$.
\label{sketch}}
\end{figure}

We now show how to understand the statistical properties of the bursts
in terms of the properties of $p(x)$.
In the following, we ignore the strongly contracting $y$ direction.
Define $\Sigma_\rho = \{(x,\rho)|\rho=\rho^* \}$ and
$\Sigma_x = \{ (x,\rho)|x=x^* \}$, and consider how the pdf $p(x)$
at $\Sigma_\rho$ evolves into a pdf $p(\rho)$ at $\Sigma_x$
(see Figure~\ref{sketch}).
When the amplitude 
$r$ becomes large (i.e. when $\rho$ becomes small), the noise terms in 
equations~(\ref{z+},\ref{z-}) are overwhelmed by the cubic deterministic 
terms.  Therefore, provided $\rho^*$ is sufficiently small, we can ignore 
the noise over the part of the trajectory shown in Figure~\ref{sketch}
and approximate the vector field by equation~(\ref{linearization}).  
The point $(x_0,\rho^*) \in \Sigma_\rho$ is mapped into 
$(x^*,f(x_0)) \in \Sigma_x$ under the flow, where 
$f(x) \equiv  \rho^* (x/x^*)^{\lambda_s/\lambda_u}$.
Given $p(x)$, the pdf $p(\rho)$ is obtained using
$|p(x) dx| = |p(\rho) d \rho|$, with $\rho = f(x)$.  For sufficiently small 
noise we can take $f(x) \approx f(\bar{x}) + f'(\bar{x}) (x-\bar{x})$,
where $\bar{x}$ is the average value of $x$ at $\Sigma_\rho$.  
Using~(\ref{pdf_x}) as the pdf for $x$ at $\Sigma_\rho$, the pdf for $\rho$ 
at $\Sigma_x$ is then
\begin{eqnarray}
p(\rho) &=& \frac{1}{\sqrt{2 \pi \sigma^2}} \exp \left[-(\rho - \bar{\rho})^2 / (2 \sigma^2) \right], 
\label{pdf_rho1} \\
\sigma^2 &=& (f'(\bar{x}))^2 \sigma_x^2, \qquad \bar{\rho} = f(\bar{x}).
\label{f_relations}
\end{eqnarray}
In particular, $p(\rho)$ is Gaussian.

\begin{figure}[t]
\begin{center}
\leavevmode
\epsfbox{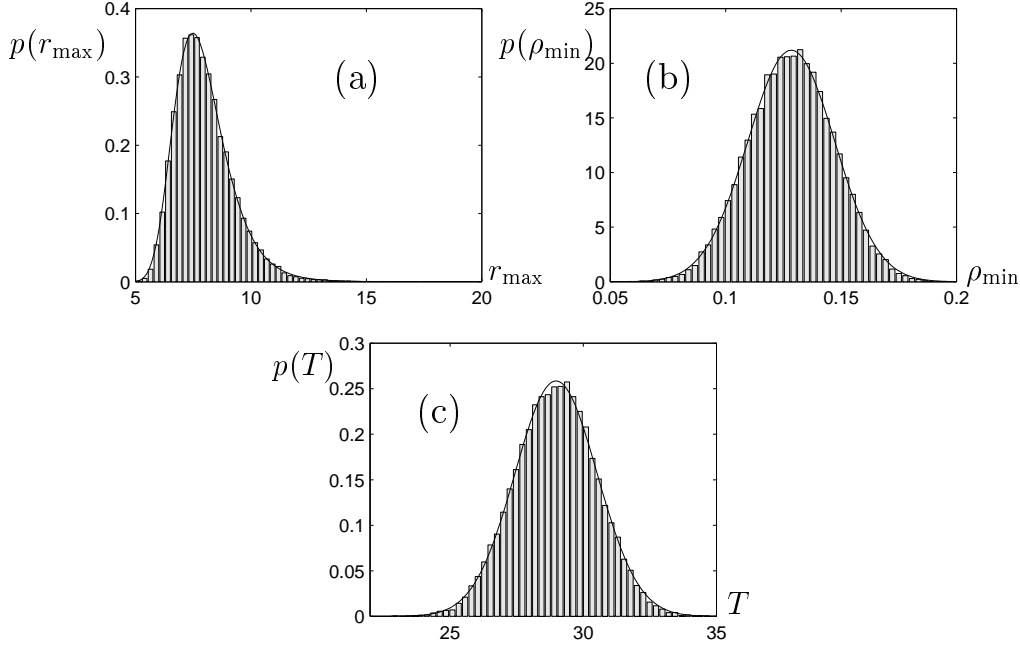}
\end{center}
\caption{
Probability distribution functions for (a) $r_{\rm max}$, (b) $\rho_{\rm min}$,
and (c) $T$ for $\lambda = 0.1, {\cal D} = 1 \times 10^{-7}$.  The 
histograms are from a long-time integration of equations~(\ref{z+},\ref{z-}), 
and the solid lines show fits given by 
equations~(\ref{pdf_rho}-\ref{pdf_T}).
\label{pdf}}
\end{figure}

\begin{figure}[t]
\begin{center}
\leavevmode
\epsfbox{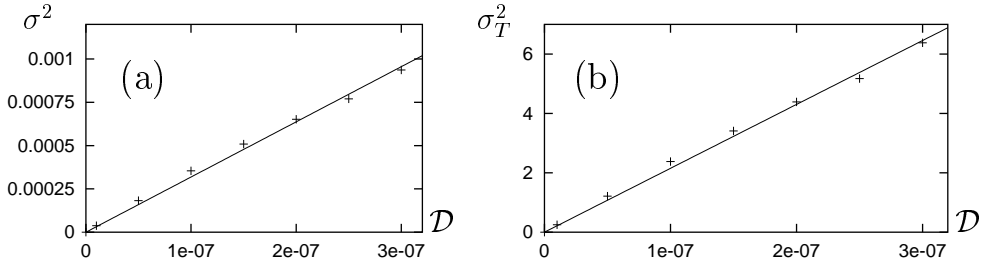}
\end{center}
\caption{
Linear dependence of the variances (a) $\sigma^2$, (b) $\sigma_T^2$
on the noise strength ${\cal D}$.
\label{pdf_fit}}
\end{figure}

Now it is assumed that in the trajectory's successive visits near 
$u_\infty$ solutions, $\rho$ always reaches its minimum value 
$\rho_{\rm min}$ at some {\it fixed} $x=x^*$.  This is reasonable because 
at some $x$ value the trajectory will have been kicked close to a $qp_\infty$ 
solution which is {\it unstable} in the $\rho$ direction, so the trajectory 
starts to evolve towards larger $\rho$ (cf.~Figure~\ref{blinking}).  With
this assumption and from~(\ref{pdf_rho1}), the pdf for $\rho_{\rm min}$ is
\begin{equation}
p(\rho_{\rm min}) = \frac{1}{\sqrt{2 \pi \sigma^2}} \exp \left[- (\rho_{\rm min} - \bar{\rho}_{\rm min})^2 / (2 \sigma^2) \right]. \label{pdf_rho}
\end{equation}
The pdf for the maximum burst amplitudes, $r_{\rm max}$, is then obtained 
using $r_{\rm max} = 1/\rho_{\rm min}$ and
$|p(r_{\rm max}) d r_{\rm max}| = |p(\rho_{\rm min}) d \rho_{\rm min}|$:
\begin{equation}
p(r_{\rm max}) = \frac{1}{r_{\rm max}^2 \sqrt{2 \pi \sigma^2}} \exp \left[-\left(\frac{1}{r_{\rm max}}-\bar{\rho}_{\rm min}\right)^2/(2 \sigma^2) \right]. \label{pdf_r}
\end{equation}
This distribution has a long $1/r_{\rm max}^2$ tail for large $r_{\rm max}$, 
indicating the probability of bursts having very large amplitudes.  
These functional forms for the pdf's are 
verified for ${\cal D} = 1 \times 10^{-7}$ in Figure~\ref{pdf}(a,b).
For the range of noise strengths considered, 
$\bar{\rho}_{\rm min} \approx 0.128$ 
independent of ${\cal D}$; this is expected from~(\ref{f_relations}) 
recalling that $\bar{x}$ was found to be independent of ${\cal D}$.  The 
variance $\sigma^2$ of the fitted distributions is linear in 
the noise strength ${\cal D}$ (see Figure~\ref{pdf_fit}(a)).
This is also expected from~(\ref{f_relations}) since $\bar{x}$
is independent of and $\sigma_x^2$ is linear in ${\cal D}$.

The statistics for the time $T$ between successive bursts are shown in 
Figure~\ref{pdf}(c) for ${\cal D}=1 \times 10^{-7}$.  The pdf for $T$ is 
fit very well by the Gaussian distribution
\begin{equation}
p(T) = \frac{1}{\sqrt{2 \pi \sigma_T^2}} \exp \left[ -(T - \bar{T})^2 / (2 \sigma_T^2) \right]. \label{pdf_T}
\end{equation}
The variance $\sigma_T^2$ of the fitted distributions is linear 
in the noise strength ${\cal D}$ (see Figure~\ref{pdf_fit}(b)), and 
$\bar{T} \approx 29.0$ is independent of~${\cal D}$.  It is interesting to 
compare these results to the results of~\cite{ston89,ston90} which
considered the effect of noise on attracting, structurally stable
heteroclinic cycles connecting {\it finite} amplitude fixed points.  It was 
found that the distribution of times between successive visits near the
fixed points on the cycle was {\it not} Gaussian; instead, it has a 
long, exponential tail for large times.  There is a fundamental reason 
for the difference between those results and the results presented here:
because the heteroclinic cycles considered in~\cite{ston89,ston90} connect 
{\it finite} amplitude solutions, the trajectory spends most its time in a 
small neighborhood these solutions.  Indeed, this is what allowed a detailed 
analysis for the times between successive visits near the fixed points to
be done.  On the other hand, the heteroclinic cycles considered in this 
paper involve {\it infinite} amplitude solutions and are traced out in 
{\it finite} time~\cite{moeh00}.  
More specifically, in the original time $t$ defined by $dt/d\tau=\rho$,
equations~(\ref{linearization}) become
\[
\frac{dx}{dt}=\frac{\lambda_u x}{\rho}, \qquad \frac{d \rho}{dt}=-\lambda_s.
\]
Assuming the initial condition $(x_0,\rho) \in \Sigma_\rho$ at $t=0$,
these equations have the solution
$x(t) = x_0 (\rho^*/(\rho^*-\lambda_s t))^{\lambda_u/\lambda_s},$ 
$\rho(t) = -\lambda_s t + \rho^*.$
The time to go from $\Sigma_\rho$ to $\Sigma_x$ is thus
$t = (\rho^*/\lambda_s) (1 - (x_0/x^*)^{\lambda_s/\lambda_u}).$
For the limit $x_0 \rightarrow 0$ (i.e., as the heteroclinic cycle is
approached), $t\rightarrow \rho^*/\lambda_s$, a {\it finite} time.
A similar argument shows that the trajectory also returns from infinity
in finite time (cf.~\cite{moeh00}).  This is why the time series in 
Figure~\ref{blinking}(a) shows very rapid growth to and decay from large 
amplitude.  We conclude that the time spent near solutions at infinity is 
not a dominant part of the total time between bursts.  Thus, the results
given in \cite{ston89,ston90} are not relevant here.

It is not possible to do detailed analytical work for the statistics
of the times between bursts because this would require a more detailed 
understanding of how the trajectory behaves away from the $u_\infty$ 
solutions.  However, the following
model for the flow between successive bursts gives some qualitative
understanding.  Consider the formal equation
\begin{equation}
\frac{d \xi}{d t} = v + \eta(t),
\label{arclength}
\end{equation}
where $v$ is constant,
and $\eta$ is a Gaussian white noise random process with 
$\langle \eta(t) \rangle = 0$ and 
$\langle \eta(t) \eta(t') \rangle = 2 {\cal D} \delta(t-t')$.  
If $\xi=\xi_0=0$ at $t=0$, the conditional probability distribution function 
$P(\xi,t|\xi_0)$ for solutions of equation~(\ref{arclength}) obeys the 
Fokker-Planck equation
\[
\partial_t P = -\partial_\xi (v P) + {\cal D} \partial_{\xi\xi} P, \qquad P = \delta(\xi) \;\; {\rm for} \;\; t=0.
\]
This has solution
\[
P(\xi,t|\xi_0=0) = \frac{1}{\sqrt{4 \pi {\cal D} t}} \exp \left(-(\xi - v t)^2/(4 {\cal D} t) \right),
\]
i.e., a Gaussian centered at $\xi=v t$ and with variance $2 {\cal D} t$.
The average time to reach a {\it fixed} value $\xi=\xi^*$ is then
$\bar{T} \equiv \xi^*/v$, and the pdf for $\xi$ at this time is 
$p(\xi) \equiv P(\xi,\bar{T}|\xi_0=0)$.  We can re-express this in terms of
a pdf for the time $T$ at which $\xi=\xi^*$ by using the deterministic
approximation $T \approx \bar{T} - (\xi-\xi^*)/v$.  Then, using
$|p(\xi) d \xi| = |p(T) dT|$, we obtain
\[
p(T) = \frac{v}{\sqrt{4 \pi {\cal D} \bar{T}}} \exp \left(-v^2 (T-\bar{T})^2/(4 {\cal D} \bar{T}) \right).
\]
We make a connection to the distribution of times between successive bursts
by thinking of $\xi$ as an arclength coordinate along a trajectory for 
equations~(\ref{z+},\ref{z-}).  Because the trajectory does not spend a
dominant part of its time in a neighborhood of the $u_\infty$ and
$qp_\infty$ solutions, equation~(\ref{arclength}) is a reasonable model
for the flow.  Assume that the arclength traced out between successive 
bursts is roughly constant.  The qualitative model~(\ref{arclength}) thus 
predicts a Gaussian distribution for the time between successive bursts, 
with the average time between bursts independent of and the variance linear 
in ${\cal D}$.  This is precisely what was found numerically.

\begin{figure}[t]
\begin{center}
\leavevmode
\epsfbox{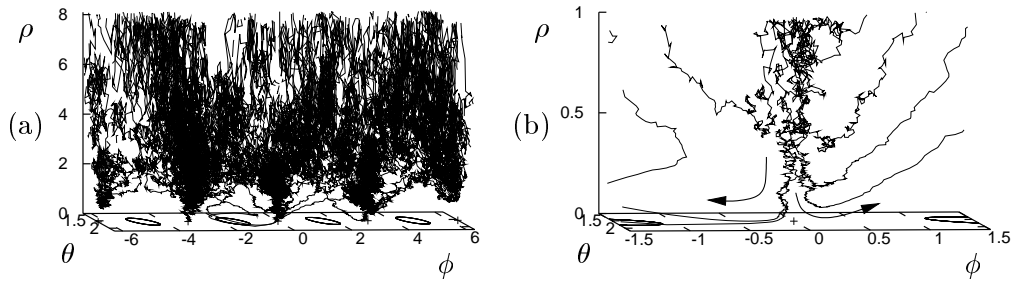}
\end{center}
\caption{(a,b) Trajectory for $\lambda = 0.1$ and ${\cal D} = 0.01$.  (b) shows
the detail near the $u_\infty$ solution at $(\rho,\theta,\phi) = (0,\pi/2,0)$,
and the arrows show the sense in which the trajectory is evolving.
\label{large_noise}}
\end{figure}

Figure~\ref{large_noise} shows the projection of a numerical solution to 
equations~(\ref{z+},\ref{z-}) with the same parameter values as for the
trajectories shown in Figure~\ref{blinking}, except with ${\cal D} = 0.01$.
Unlike the result in Figure~\ref{blinking}(b) for much smaller noise 
strength, here the underlying stable quasiperiodic solution is not even 
recognizable.  The arrows in Figure~\ref{large_noise}(b) show
that the trajectory can make a visit near a $u_\infty$ solution and depart 
either to the ``left'' or to the ``right''.  In the $(\rho,x,y)$ coordinates
introduced above (cf. equation~(\ref{linearization})), the trajectory
can intersect a surface of constant $\rho$ for either positive or negative
$x$ values.  Thus, sufficiently large noise can be responsible for 
the trajectory crossing the stable manifold of the $u_\infty$ solutions.
This leads to the possibility that successive visits near infinity
can either be to {\it different} or to the {\it same} $u_\infty$ solutions.
In the convection system, successive visits near the {\it same} $u_\infty$ 
solution correspond to successive bursts occurring on the {\it same} side 
of the container~\cite{moeh98}.  Therefore, sufficiently
large noise can affect the physical manifestation of the bursts by
destroying purely blinking states.

\begin{figure}[t]
\begin{center}
\leavevmode
\epsfbox{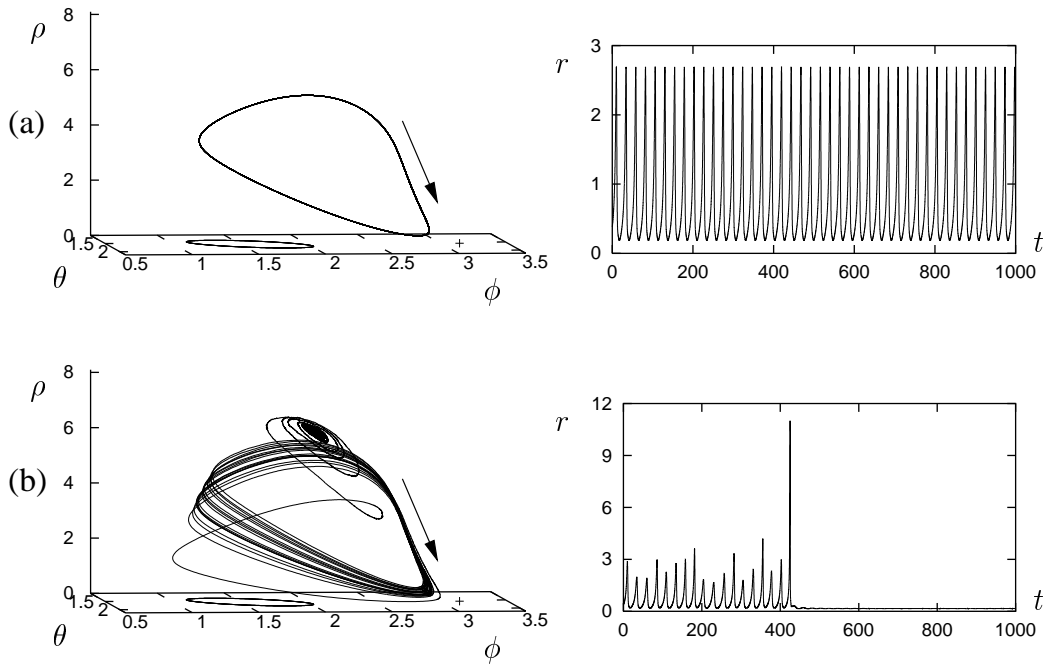}
\end{center}
\caption{
Bursts for $\lambda = 0.1253$ and (a) ${\cal D} = 0$, (b) 
${\cal D} = 1 \times 10^{-7}$.  In (b), it is seen that the noise
kicks the system out of the basin of attraction of the state shown
in (a) and into the basin of attraction of a solution with constant 
$\rho=1/r$.
\label{winking}}
\end{figure}

Reference~\cite{moeh98} also identified a different type of stable
quasiperiodic solution which is present in the absence of noise for
$\lambda = 0.1253$ (in the notation of~\cite{moeh00}, this is a $u/v^1$
solution).  The trajectory for this solution makes successive
visits always near the {\it same} $u_\infty$ solution (see 
Figure~\ref{winking}(a)).
Physically, this corresponds to a {\it winking state} for the convection
system because successive bursts occur at the same side of the 
container~\cite{moeh98}.  This solution coexists with stable, finite
amplitude periodic solutions at 
$(\rho,\theta,\phi) = (6.298,1.753,2.010\pm m \pi)$, where $m$ is an 
integer.  Reference~\cite{moeh98} argues that these stable, finite
amplitude periodic solutions are more likely to be observed in experiments
in which the Rayleigh number is ramped upwards, which helps to explain
why winking states have apparently never been observed in binary fluid
convection experiments.  It is seen that even a very small amount of noise 
$({\cal D}=1 \times 10^{-7})$ can kick the system out of the basin
of attraction of the winking state and into the basin of attraction of
a stable periodic solution (see Figure~\ref{winking}(b)).  Thus, even
if a winking state could be established in an experiment, it would be
particularly sensitive to noise.

\section{Noise-induced bursting}
\begin{figure}[t]
\begin{center}
\leavevmode
\epsfbox{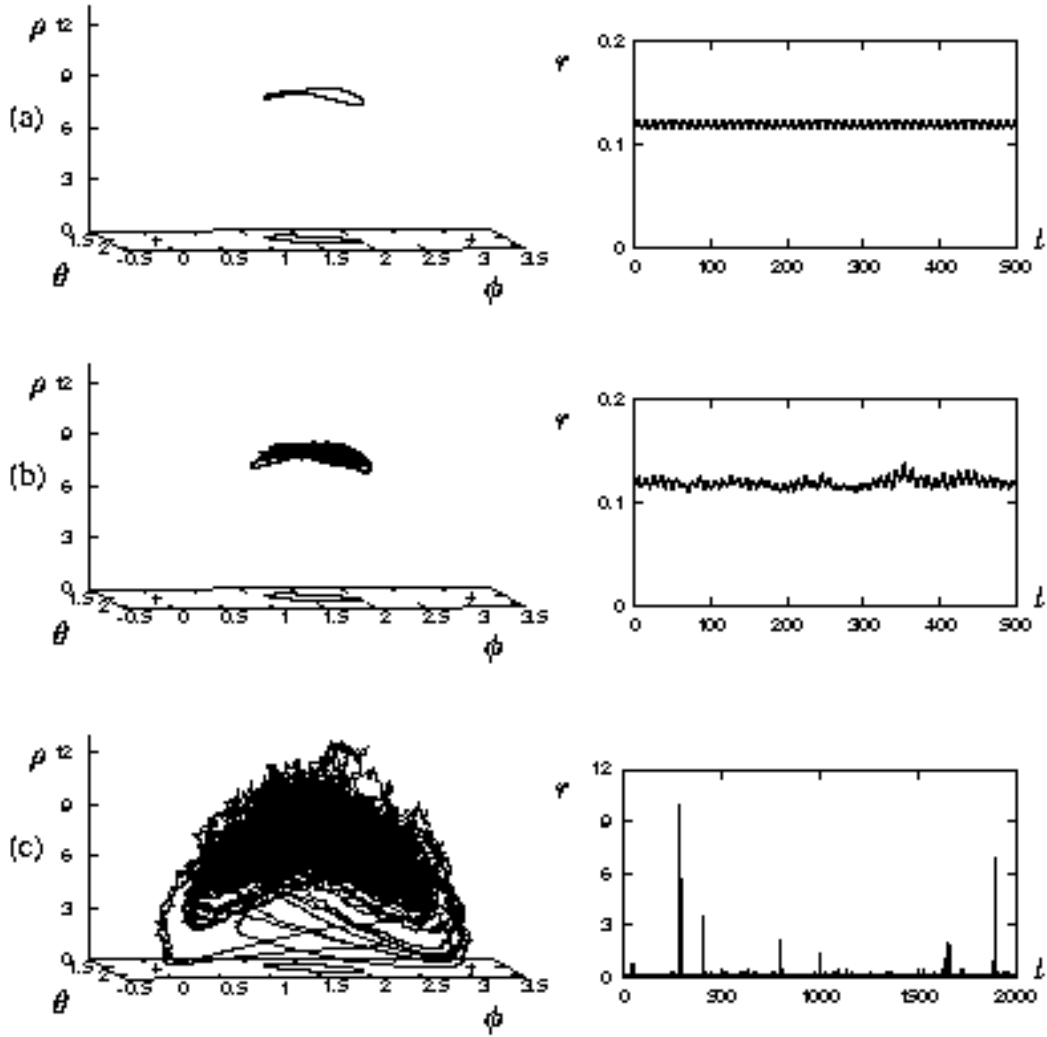}
\end{center}
\caption{
Solutions for $\lambda = 0.1$ and (a) ${\cal D} = 0$, (b) 
${\cal D} = 1 \times 10^{-6}$, (c) ${\cal D} = 1 \times 10^{-4}$ 
with $\Delta \lambda = \Delta \omega = 0$. 
Note the different scales for the time series plots.
\label{noise_induced}}
\end{figure}

It will now be shown that related bursts may occur for 
equations~(\ref{z+},\ref{z-}) even in the absence of forced 
symmetry-breaking, that is, when noise is added to the normal form 
equations for the Hopf bifurcation with {\it exact} D$_4$ symmetry.  
Specifically, consider equations~(\ref{z+},\ref{z-}) with the same
parameters as considered in the previous section
($A=1-1.5 i, B=-2.8+5 i, C=1+i,\omega=1$), except with 
$\Delta \lambda=\Delta \omega=0$.  For these coefficient values in
the absence of noise, a branch of unstable periodic solutions bifurcates 
subcritically from the trivial state, and two branches of unstable periodic
orbits and one branch of globally attracting quasiperiodic solutions all 
bifurcate supercritically~(see \cite{swif88}, also~\cite{moeh98,moeh00}).  
Infinite amplitude periodic and quasiperiodic solutions exist as
counterparts of the finite amplitude solutions which bifurcate from the
trivial state.  In particular, $u_\infty$ and $qp_\infty$ solutions are the 
infinite amplitude counterparts of the periodic solutions on the subcritical 
branch and quasiperiodic solutions on the supercritical branch, 
respectively~\cite{moeh00}.  
If for some reason the trajectory comes close to the stable manifold of a 
$u_\infty$ solution, it will make an excursion to a neighborhood of that 
solution, then get kicked toward a $qp_\infty$ solution which returns it 
to smaller amplitude.  However, since the finite amplitude quasiperiodic 
solutions are {\it globally} attracting, heteroclinic cycles involving 
infinite amplitude solutions cannot form, so such a burst is a 
transient phenomenon.  A stable quasiperiodic solution is shown for 
$\lambda = 0.1$, ${\cal D} = 0$ in Figure~\ref{noise_induced}(a).  
For small noise the trajectory remains near the 
quasiperiodic solution (see Figure~\ref{noise_induced}(b)).  However, 
sufficiently large noise can repeatedly kick the system close to the 
stable manifold of a $u_\infty$ solution, leading to bursts of very large 
dynamic range (see Figure~\ref{noise_induced}(c)).  This is highly
reminiscent of results for the effect of resonant temporal forcing on the 
Hopf bifurcation with D$_4$ symmetry~\cite{moeh00b}; there it was found
that as the forcing amplitude increased, the attractor came closer to the
stable manifolds of $u_\infty$ solutions, leading to bursting behavior.

\begin{figure}[t]
\begin{center}
\leavevmode
\epsfbox{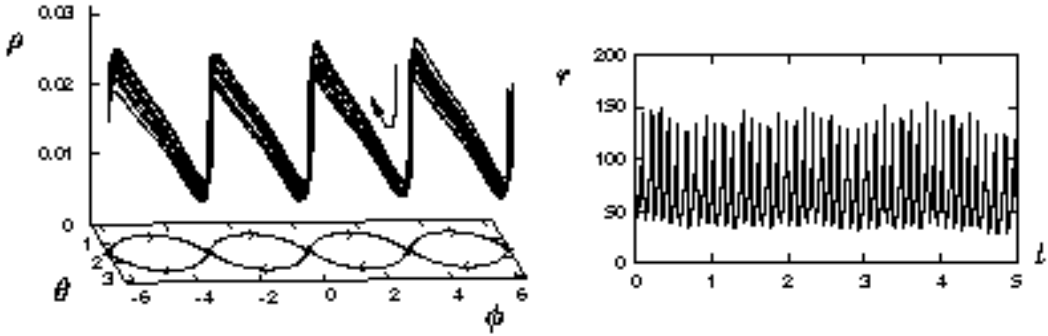}
\end{center}
\caption{The left figure shows the attracting infinite amplitude 
heteroclinic cycle connecting $u_\infty$ solutions at $\rho=0$, and
the ``statistical limit cycle'' solution for ${\cal D} = 1 \times 10^{-5}$.
The right figure shows that successive visits near $u_\infty$ solutions
occurs randomly, but with a well-defined mean period.
\label{hetero}}
\end{figure}

As a further demonstration of the effect of noise on a Hopf bifurcation with 
exact D$_4$ symmetry, we close by considering an {\it attracting} infinite 
amplitude heteroclinic cycle.  Specifically, consider the parameters 
$A = 0.1 - 1.5 i, B = -1 - 4 i, C = 1 - 2 i, \lambda = 0.1,
\omega = 1, \Delta \lambda = \Delta \omega = 0$.  Using the results 
of~\cite{swif88}, it can be shown that there are heteroclinic 
connections between $u_\infty$ solutions {\it within} the $\Sigma$ 
subspace.  Linearizing equations~(\ref{rho}-\ref{phi_rho}) about the $u_\infty$
solutions shows that the eigenvalue corresponding to perturbations
in the $\rho$ direction is $-0.2$, while the eigenvalues corresponding
to perturbations within the $\Sigma$ subspace are $-4.828$ and $0.828$.
Since the first eigenvalue is negative and the sum of the latter two 
eigenvalues is also negative, the heteroclinic cycle connecting $u_\infty$
is {\it attracting} (cf.~\cite{swif88}).  Figure~\ref{hetero} shows a
numerically calculated trajectory for these parameter values and
${\cal D} = 1 \times 10^{-5}$; the time step of integration is 
$\delta t = 1 \times 10^{-6}$.  
Much like the situations described in~\cite{buss80,ston89,ston90}, noise 
causes the attracting heteroclinic cycle to become a ``statistical limit 
cycle'' in which switching between $u_\infty$ solutions occurs randomly in 
time but with a well-defined mean period.

\section{Conclusion}
The effect of additive white noise on a model for large aspect-ratio binary 
fluid convection has been considered.  Particular attention has been paid
to the effect of noise on bursting behavior present in the model.
It was shown that even a very small amount of noise can have a very large 
effect on the amplitudes of successive bursts.  This is because there will 
be a burst with larger amplitude if the noise kicks the system closer to 
the stable manifold of an infinite amplitude state, and a burst with smaller 
amplitude if the noise kicks it away.  Numerical results and analytical 
arguments were given to understand the statistical properties of bursts 
in the presence of noise.  It was also demonstrated that large enough noise 
can affect the physical manifestations of the bursts by destroying purely 
blinking and purely winking states.  In fact, winking states are 
particularly sensitive to noise, which helps to explain why they have
apparently never been observed in experiments.  Finally, it was shown 
that related bursts can occur when noise is added to the normal form 
equations for the Hopf bifurcation with {\it exact} square symmetry.

Reasoning similar to that given in~\cite{moeh00} suggests that this type of
bursting behavior can persist when higher terms in 
equations~(\ref{z+},\ref{z-}) are retained, even in the presence of noise.  
Bursts will then be associated with visits near large but {\it finite} 
amplitude solutions.  This is important because it implies that bursts 
similar to those described in this paper may be observed in real physical 
systems undergoing a Hopf bifurcation with exact or weakly broken D$_4$ 
symmetry; the (unphysical) infinite amplitude solutions are not necessary.
There is also a possibility that in appropriate cases the noise could be 
more important than higher order terms so their exact nature would be
irrelevant.

Other physical systems for which the results presented in the paper might
be relevant because their evolution equations are related to the Hopf
bifurcation with D$_4$ symmetry include any system in a square domain 
undergoing an oscillatory instability~\cite{ashw95} or displaying
oscillatory patterns which are periodic on a square 
lattice~\cite{silb91}, electrohydrodynamic convection in liquid 
crystals~\cite{riec94}, lasers~\cite{feng94}, spring-supported
fluid-conveying tubes~\cite{stei95}, the Faraday system in a square
or nearly-square container~\cite{simo89}, and dynamo theories of magnetic
field generation in the Sun~\cite{knob96b,knob98b}.

{\bf Acknowledgements}:
This work was supported by a National Science Foundation Mathematical
Sciences Postdoctoral Research Fellowship.  I would like to thank Edgar 
Knobloch and Philip Holmes for useful comments and discussions.

\end{document}